\documentclass[prb, twocolumn ,reprint]{revtex4-1}

\usepackage{epsfig}
\usepackage{amsmath}
\usepackage{amssymb}
\usepackage{mathrsfs}
\usepackage{mathtools}
\usepackage{comment}
\usepackage{color}
\usepackage{braket}
\usepackage{nicefrac}
\usepackage{yfonts}
\usepackage{bm}
\usepackage{longtable}
\usepackage{enumerate}
\newcommand{\be}{\begin{equation}}
\newcommand{\ee}{\end{equation}}
\newcommand{\ba}{\begin{aligned}}
\newcommand{\ea}{\end{aligned}}

\newcommand{\btr}[2]{\mathrm{Tr}_{#1}\bigl[#2\bigr]}

\newcommand{\bea}{\begin{eqnarray}}
\newcommand{\eea}{\end{eqnarray}}

\begin{document}
\title{Conservation laws for a class of generic Hamiltonians}
\author{Maurizio Fagotti$^{1}$}
\affiliation{$^1$The Rudolf Peierls Centre for Theoretical Physics, Oxford University, Oxford, OX1 3NP, United Kingdom}
\begin{abstract}
Within a strong coupling expansion, we construct local quasi-conserved operators for a class of Hamiltonians that includes both integrable and non-integrable models.  
We explicitly show that at the lowest orders of perturbation theory the structure of the operators is independent of the system details. 
Higher order contributions are investigated numerically by means of an \emph{ab initio} method for computing the time evolution of local operators in the Heisenberg picture. The numerical analysis suggests that the quasi-conserved operators could be approximations of a quasi-local conservation law, even if the model is non-integrable.   
\end{abstract}
\maketitle
The importance of conservation laws in physics can not be overestimated.  In quantum mechanics conserved quantities are fundamental for many reasons, starting from the classification of the states (quantum numbers),  the one-to-one correspondence with the symmetries of the system (Noether's theorem), the key role played in ergodicity and transport phenomena\cite{ergo,mazur, suzuki}, the issue of integrability\cite{ABA}.

In the last decade conservation laws have been in the spotlight of non-equilibrium quantum many-body physics in low dimensional systems for their influence on the relaxation of local degrees of freedom. 
Local conservation laws have been identified\cite{GGE,thermalisation, thermalisation1,thermalisation2,thermalisation3, thermalisation4, RDMGGE} as the basic elements that characterise the stationary behaviour of local observables at late time after a sudden change of a global Hamiltonian parameter (global quench). 
This picture was corroborated by many analytical and numerical results\cite{GGEIsing, GGEdyn, RDMGGE,RelaxXXZ, cluster,velXXZ, Bosegas, localrelax, centrallimit, QFT, Luttinger}, which led to a better understanding of the type of conservation laws that are relevant for the non-equilibrium problem. Nevertheless, recent works\cite{quenchaction,quenchactionMG, Andrei} have been pointing up some inconsistencies that might potentially undermine the established physical picture.  

In this kind of situation, an intimate knowledge of the models under investigation is essential: disregarding a single conservation law may lead to the wrong conclusion\cite{RDMGGE, nonabelian}. 

For example, in the absence of manifest symmetries, it is widely believed that generic models do not possess (quasi-)local conservation laws. The hypothesis of local thermalisation\cite{thermalisation,thermalisation1,thermalisation2,thermalisation3,thermalisation4} is a crucial result based on that assumption. 
A non-integrable model could however be not as generic as one would expect\cite{ErgPros, absenceT, thermnat},
resulting in behaviours difficult to understand (\emph{cf}. \cite{strong-weakT}). 

In integrable models the situation is not clearer. By definition there is an infinite number of local conservation laws that account for integrability\cite{ABA}, but other infinitely many (quasi-)local charges could be present as a result of extra symmetries\cite{nonabelian, qlcons,qlcons1,qlcons2}. In addition, the set of conservation laws could be ``oversized'' and non-abelian\cite{nonabelian}.

In this paper we investigate a class of spin chain Hamiltonians that include many well-known integrable and non-integrable models.
We carry out a strong coupling expansion and construct local operators that are quasi-conserved. 

We conjecture that the quasi-conserved operators are an approximation of a conservation law, which can  be formally written without knowing the system details. 
To the best of our knowledge, this has never been pointed out in its full generality until now.
 
Finally, we investigate the locality properties of the conservation law. 

\paragraph{Locality.}
In a spin chain we call `local' a translation invariant operator $\mathcal O$ that can be written as
\be\label{eq:ql}
\mathcal O=\sum_\ell  \hat o_\ell\, ,
\ee
where $\hat o_\ell$ are operators that act like the identity everywhere but on a finite number of sites around $\ell$.   The \emph{range} of \eqref{eq:ql} is defined as the minimal length of the interval on which $o_\ell$ acts non-trivially. We notice that the commutator of local operators is local.

Having in mind the non-equilibrium problem\cite{Note1}, we say that $\mathcal O$ is `weakly local' if 
\be\label{eq:norm}
\lim_{|A|\rightarrow\infty}\parallel \hat o_\ell-\hat o_\ell^{(A)}\parallel=0\, ,
\ee
where $\parallel\cdot\parallel$ is the operator norm (the maximal eigenvalue in absolute value), $A$ is a  region surrounding $\ell$, $|A|$ its length, and
\be\label{eq:OA}
\hat o_\ell^{(A)}=\frac{\btr{\bar A}{\hat o_\ell}}{\btr{\bar A}{\mathrm I_{\bar A}}}\otimes \mathrm I_{\bar A}
\ee
is the operator restricted to $A$ ($\bar A$ is complementary to $A$ and $\mathrm I_{\bar A}$ is the identity on $\bar A$).

We also remind the reader of the definition of \emph{quasi-locality} (see \emph{e.g.} \cite{exploc}), which is a stronger form of \eqref{eq:norm} with the additional requirement of \emph{exponential localisation} \mbox{$\parallel \hat o_\ell-\hat o_\ell^{(A)}\parallel\sim e^{-|A|/r}$}, where $r$ is the \emph{typical range}. 

\paragraph{The class of models.}
\begin{table}
\begin{tabular}{c|c|c}
model&$H_I$&$H_0$\\
\hline
XYZ${}^{(1)}$&$\frac{1}{4}\{zz\}$&$J_x\{xx\}+J_y\{yy\}+\sum\limits_{\alpha}h_\alpha\{\alpha\}$\\
XYZ${}^{(2)}$&$\frac{1}{4}\{z\}$&$J_x\{xx\}+J_y\{yy\}+J_z\{zz\}$\\
Ising&$\frac{1}{2}\{z\}$&$J_x\{zz\}+h_x\{x\}+h_y\{y\}$\\
ANNNI${}^{(1)}$&$\frac{1}{4}\{x\}$&$J\{zz\}+g\{z0 z\}$\\
ANNNI${}^{(2)}$&$\frac{1}{4}\{zz\}$&$g\{z0 z\}+h_x\{x\}+h_y\{y\}$\\
ANNNI${}^{(3)}$&$\frac{1}{4}\{z0 z\}$&$J\{zz\}+h_x\{x\}+h_y\{y\}$
\end{tabular}
\caption{A list of Hamiltonians with the property \eqref{eq:algebra}. They describe both integrable and non-integrable models. 
Notations: 
$\{\alpha\beta\hdots\}=\sum_\ell \sigma_\ell^\alpha  \sigma_{\ell+1}^\beta\cdots$,
where, for $\alpha\in\{x,y,z\}$,  $\sigma^\alpha_\ell$ acts like the corresponding Pauli matrix on site $\ell $ and like the identity elsewhere, while $\sigma^0_\ell$ means identity.  }\label{t:1}
\end{table} 

We consider local translation invariant Hamiltonians of the form
\be\label{eq:H0}
H=H_0+\Delta H_I\, ,
\ee
where $H_0=G+F+F^\dag$ and $G$ and $F$ are local operators satisfying the algebra
\be\label{eq:algebra}
[H_I,G]=0\, ,\qquad [H_I,F]=F\, .
\ee 
This class includes widely studied models, some of which are reported in Table \ref{t:1} (in particular, the algebraic structure \eqref{eq:algebra} was used in \cite{Hubbard} to compute a $t/U$ expansion for the Hubbard model). 
We notice that for some models the parametrisation \eqref{eq:H0}\eqref{eq:algebra} is not unique but depends on the choice of $H_I$.   

\paragraph{Quasi-conserved operators.}
For large $\Delta$ we consider operators that can be series expanded in powers of $\Delta^{-1}$
\be
Q=\sum_{n=0}^\infty \frac{Q_n}{\Delta^n} \, .
\ee  
We implicitly assume that the expansion makes sense. 
Clearly $H/\Delta$ is an operator of that kind. The commutator $[H,Q]$ can be series expanded as well
\be
[H,Q]=\Delta [H_I,Q_0]+\sum_{n=0}^\infty \frac{[H_0,Q_n]+[H_I,Q_{n+1}]}{\Delta^n} \, .
\ee
We say that $Q^{(\alpha)}$ is quasi-conserved at order $O(\Delta^{-\alpha})$ if the following system is satisfied:
\be\label{eq:eqcons}
\left\{
\ba
&[H_I,Q^{(\alpha)}_0]=0\\
&[H_0,Q^{(\alpha)}_n]+[H_I,Q^{(\alpha)}_{n+1}]=0\qquad n\leq \alpha\, .
\ea
\right.
\ee
In particular, we can set $Q^{(\alpha)}_{n}=0$ for $n>\alpha+1$. 

The construction of conserved quantities by working out \eqref{eq:eqcons} is generally unfeasible and indeed in integrable models more powerful techniques have been developed\cite{ABA}.  
We are however interested in generic models, so we can only rely on the algebra \eqref{eq:algebra}. 

We use \eqref{eq:algebra} to ``integrate'' $H_I$ out of the system of equations \eqref{eq:eqcons}. From \eqref{eq:algebra} and \eqref{eq:eqcons} it follows that the operator
$Q^{(\alpha)}_n(t)=e^{i H_I t} Q^{(\alpha)}_n e^{-i H_I t}$  is $2\pi$-periodic in $t$ and can therefore be expanded as a Fourier series 
\be\label{eq:QOmega}
Q^{(\alpha)}_n(t)=\sum_{j}e^{i j t}\Omega_{n,j}\, ,
\ee
where $\Omega_{n,-j}=\Omega^\dag_{n,j}$. 
The problem is now reduced to the calculation of $\Omega_{n,j}$.
To shorten the notations, we introduce an auxiliary parameter $s$ and consider a more general class of operators 
\be\label{eq:def}
\Omega_{n,j}(s)=e^{i G s} \Omega_{n,j} e^{-i G s}\, ,\quad F(s)=e^{i G s} Fe^{-i G s}\, .
\ee 
Using ansatz \eqref{eq:QOmega}, the system of equations \eqref{eq:eqcons} for $\Omega_{n,j}(s)$ reads as
\be\label{eq:sys}
\left\{
\ba
\dot\Omega_{0,0}&=0\\
\Omega_{n,j}&=0\qquad |j|>n\\
\Omega_{n+1,j}&=\frac{i \dot \Omega_{n,j}+[\Omega_{n,j-1},F]+[\Omega_{n,j+1},F^\dag]}{j}\\
i\dot \Omega_{n,0}&=[F,\Omega_{n,-1}]+[F^\dag,\Omega_{n,1}]\, ,\\
\ea
\right.
\ee
where all the operators are functions of $s$ and the dot is used to indicate the derivative with respect to $s$. 
$Q^{(\alpha)}$ is obtained going backwards through the various steps: rewriting derivatives in terms of commutators ($\dot{\mathcal O}=i[G,\mathcal O]$), setting $s=0$, and summing $\Omega_{n,j}$ over j (\emph{cf}. \eqref{eq:QOmega} with $t=0$). 

The Hamiltonian is generated by  $\Omega_{0,0}=H_I$, using then \eqref{eq:algebra} to remove $H_I$ from the equations.

\paragraph{A formal conservation law.}

The first equation of \eqref{eq:sys} has another formal solution: $\Omega_{0,0}=G$. Let us attempt to generate quasi-conserved operators that approach $G$ as $\Delta\rightarrow\infty $. At fixed order $\alpha$, we seek a solution that can be written in terms of $F(s)$ and a finite number of its derivatives. The latter condition is sufficient  for locality. 

The only obstacle to achieving that goal is the inversion of the last equation of \eqref{eq:sys}.
If, order by order, we are able to find a primitive of $([F^\dag(s),\Omega_{n,1}(s)]-{\rm h.c.})$, then \eqref{eq:sys} can be  solved recursively. 

Before going any further, let us notice that we can always add to $\Omega_{n,0}$ a term proportional to $G$ or $H_I$ (\emph{cf}. \eqref{eq:sys}, $\dot G=\dot H_I=0$). However, since the equations are linear in $\Omega$, this would correspond to multiplying the full operator by a constant or to adding a term proportional to the Hamiltonian. Thus, we are going to ignore these degrees of freedom. 

In Table \ref{t:2} the solution of the system up to $O(\Delta^{-4})$ is reported. 
The expressions become quickly cumbersome as the order of the approximation is increased, however we stress that everything is written in terms of (nested) commutators and therefore the operators remain local. 

More generally, we conjecture that $\Omega_{n,0}$ can be determined unequivocally\cite{SM}. 

The possibility of expressing the lowest orders of perturbation theory in closed form (\emph{cf.} Table \ref{t:2}) is a strong indication that $Q^{(\alpha)}$ are approximations of a conservation law $Q$ that can be formally written without knowing the system details. 
From this point of view, $Q$ is the unique formal solution of \eqref{eq:sys} independent of $H$ for given $H_I$ (redefining $H_I$ other solutions can be found).

\begin{widetext}
\begin{longtable}{c||c|c|c|c}
\caption{The lowest orders of the perturbative expansion. For $n=4$ we only report $\Omega_{4,0}$, which is the only operator that can not be obtained with mere recursion. The shorthands  in the last row stand for the nested commutators obtained by placing  $F$, $F$, $F^\dag$, and $F^\dag$ (in this exact order) between commas; the dot means derivative; \emph{e.g.} $[,[,[,\cdot]]]\equiv [F,[F,[F^\dag,\dot F^\dag]]]$. }\label{t:2}\\
$n$&$\Omega_{n,0}$&$\Omega_{n,1}$&$\Omega_{n,2}$&$\Omega_{n,3}$\\
\hline
\hline
0&$G$&$0$&$0$&$0$\\
1&$[F,F^\dag]$&$-i \dot F$&$0$&$0$\\
2&$i([\dot F,F^\dag]-[F,\dot F^\dag])$&$\ddot F-[F,[F,F^\dag]]$&$\frac{i}{2} [F,\dot F]$&$0$\\
3&$[\dot F,\dot F^\dag]-[\ddot F,F^\dag]-[ F,\ddot F^\dag]+\frac{3}{2}[[F,[F,F^\dag]],F^\dag]$&$i\dddot F-3i[F,[\dot F,F^\dag]]+\frac{3 i}{2}[[ F, \dot F],F^\dag]$&$\frac{1}{2}[F,[F,[F,F^\dag]]]+\frac{3}{4}[\ddot F,F]$&$\frac{i}{6}[F,[\dot F,F]]$\\[0.1mm]
\cline{3-5}\\[0.1mm]
4 &\multicolumn{4}{c}{
$i([F,\dddot F^\dag]-[\dot F,\ddot F^\dag]+[\ddot F,\dot F^\dag]-[\dddot F,F^\dag]+\frac{9}{4}[,[,[\cdot,]]]-\frac{9}{4}[[[,\cdot],],]-4[,[[,\cdot],]]+4[[,[\cdot,]],]-[,[[\cdot,],]]+[[,[,\cdot]],])$
}\\
\end{longtable}
\end{widetext}

Incidentally, our approach might call to mind the Schrieffer-Wolff transformation\cite{Hubbard,SWtransf0,SWtransf}.  
However, it does not seem that $Q$ can be related to the Schrieffer-Wolff effective Hamiltonian in a trivial way. 

\paragraph{Locality properties.}
In order to gain some insights into the locality properties of $Q$, let us restore the dependence on $\Delta$. The system of equations \eqref{eq:sys} is such that a factor $\Delta^{-1}$ appears for each operator $F$ and for each derivative (\emph{i.e.}, for each $G$). 
We therefore expect a solution of the form
\be
Q=G+\Delta \mathcal F(\hat f_\Delta)\, ,
\ee
where $ \mathcal F$ is a functional independent of $\Delta$ and 
\be\label{eq:fD}
\hat f_\Delta(s)=\frac{F(s/\Delta)}{\Delta}
\ee
is a quasi-local operator with typical range proportional to $s/\Delta$ (\emph{cf.} \eqref{eq:def}, the proportionality constant is essentially given by the Lieb-Robinson velocity associated with $G$ \cite{localbounds}).
The larger $\Delta$ is, the smaller is the size of the region around $s=0$ where the behaviour of $\hat f_\Delta(s)$ significantly affects the conservation law. In addition, because of the overall factor $\Delta^{-1}$ in \eqref{eq:fD}, the 
contribution of nested commutators of $\hat f_\Delta(s)$ in $\mathcal F$ is exponentially suppressed with the number of operators involved. 
Because each commutator increases the typical range of the operator by a finite amount, one can naively expect that $Q$ is a quasi-local conservation law with a typical range that scales as $1/\log(\Delta/\Delta_0)$, where $\Delta_0$ depends on the system details.   

In addition, some preliminary results (see also \cite{SM}) show that in integrable models $Q$ is closely related to the operator obtained from the Hamiltonian by flattening the excitation energies, sending any interaction parameter different from $\Delta$ (\emph{e.g.} $g$, $h$, and $J$ in Table \ref{t:1})  to zero. In non-interacting models the latter condition is sufficient for quasi-locality, however in the presence of interactions and especially in generic models the statement of quasi-locality can only be accepted with reserve.

We partially handle this weakness by studying the range of the quasi-conserved operators numerically. 

\paragraph{Time averaged operators.}
Let us consider the time average $\bar{\mathcal O}$ of an operator $\mathcal O$ in the Heisenberg picture 
\be
\bar{\mathcal O}(t)=\int_0^t\frac{\mathrm d \tau}{t}\mathcal O(\tau)\, ,\quad \mathcal O(\tau)=e^{i H \tau} \mathcal O e^{-i H \tau}\, .
\ee
If $\mathcal O$ is conserved, $\bar{\mathcal O}$ is independent of time and equal to $\mathcal O$. More generally, at large time the time average could approach zero, so it could be convenient to define the operator $\bar {\mathcal O}_{\ast}$, normalised according to a given norm
\be\label{eq:normalization}
\bar{\mathcal O}_{\ast}(t)=\bar{\mathcal O}(t)/\parallel \bar{\mathcal O}(t)\parallel_\ast\, .
\ee
The norm of the commutator between $\bar{\mathcal O}_{\ast}(t)$ and the Hamiltonian is given by
\be\label{eq:tdecay}
\parallel [H,\bar{\mathcal O}_{\ast}(t)] \parallel_\ast= \frac{\parallel \mathcal O(t)-\mathcal O\parallel_\ast}{t \parallel \bar{\mathcal O}(t)\parallel_\ast}\leq \frac{2 \parallel \mathcal O\parallel_\ast
}{t \parallel \bar{\mathcal O}(t)\parallel_\ast}\, .
\ee
If $\lim_{t\rightarrow\infty }t \parallel \bar{\mathcal O}(t)\parallel_\ast = \infty$,  the operator $\bar{\mathcal O}_{\ast}(t)$ is quasi-conserved at fixed (sufficiently large) time and conserved as $t\rightarrow\infty$. Because of the Lieb-Robinson bounds\cite{localbounds}, $\bar{\mathcal O}_{\ast}(t)$ is a quasi-local operator.   
For generic local $\mathcal O$, the typical range of $\bar{\mathcal O}_{\ast}(t)$ could however increase in time, making $\bar{\mathcal O}_{\ast}(t)$ nonlocal as $t\rightarrow\infty$.

\paragraph{Numerical method.} 
We use a simple algorithm to compute the time average of $Q^{(\alpha)}(t)$ \eqref{eq:eqcons} in the thermodynamic limit. It is based on the fact that commutators of translation invariant operators in spin chains can be computed efficiently. We implemented the time evolution in the Heisenberg picture by discretising the time and solving the following equation (which is exact for any $\delta t_n=t_n-t_{n-1}$)
\be
\bar{\mathcal O}(t_n)=\Bigl(1-\frac{\delta t_n}{t_n}\Bigr)e^{i\mathcal H\delta t_n}(\bar{\mathcal O}(t_{n-1}))+\frac{\delta t_n}{t_n} e^{i\mathcal H\delta t_n} (\mathcal O)\, ,
\ee
where $\mathcal H$ is the superoperator associated with the time evolution $\mathcal H(\mathcal O)=[H,\mathcal O]$. In practice, we expanded $e^{i\mathcal H\delta t}$ at the second order in $\delta t$ and set a lower bound (inversely proportional to the time) to the norm of the operators retained. 
Since the algorithm breaks unitarity, the norm of the time evolving operator is an important parameter to keep under control.   

Let us focus on non-integrable models.
If the quasi-conserved operators $Q^{(\alpha)}$ are approximations of a quasi-local conservation law $Q$, one might expect 
\be\label{eq:exp}
\tilde Q^{(\alpha)}(t)\xrightarrow{t\rightarrow \infty} \gamma  Q+\beta H \quad(+\hdots)\, ,
\ee 
where $\tilde Q^{(\alpha)}$ is the numerically computed time-average. 

As a matter of fact, we can not rule out the presence of other conservation laws, also nonlocal, on the right hand side of \eqref{eq:exp}.  
This could make us believe that $Q$ is nonlocal even if it is not. Nevertheless, in the following we are going to assume \eqref{eq:exp}.

\paragraph{Quantum Ising model.} As an explicit example we consider the quantum Ising model in a magnetic field with nonzero transverse and longitudinal components
\be\label{eq:Is}
H=\sum_\ell\Bigl(\frac{1}{4}\sigma_\ell^z\sigma_{\ell+1}^z+\frac{g}{2} \sigma_\ell^x+\frac{\Delta}{2}\sigma_\ell^z\Bigr)\, .
\ee
For generic $g$ and $\Delta$ this describes a non-integrable model. It has however some nice properties (\emph{e.g.}  $F(s)$ is local for any $s$) that help reducing the numerical effort of the analysis. 

\begin{figure}[tbp]
\includegraphics[width=0.45\textwidth]{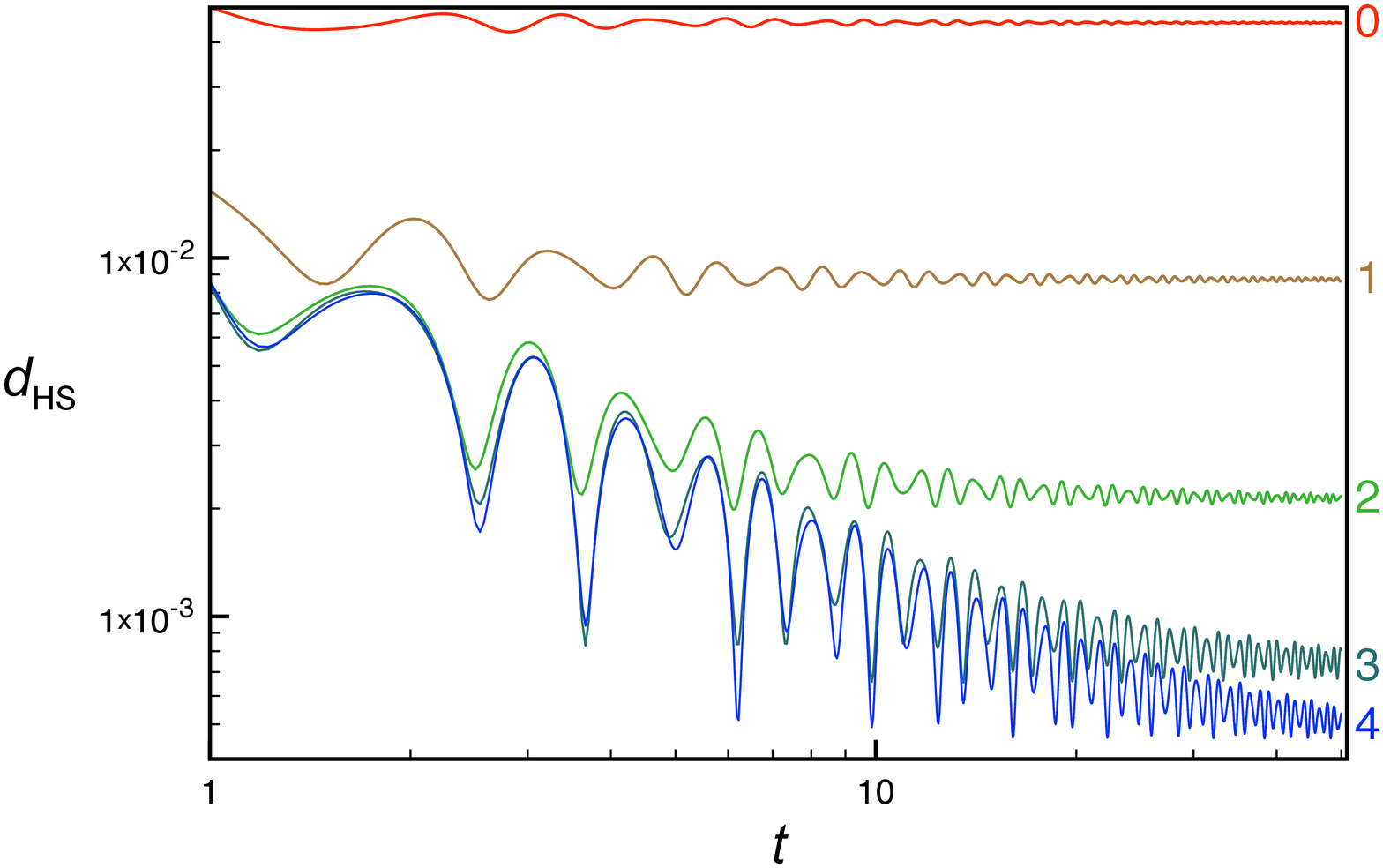}
\caption{The Hilbert-Schmidt distance between the (opportunely normalised, see \eqref{eq:dHS} of \cite{SM}) time average $\tilde G(t)$ and the quasi-conserved operators $Q^{(\alpha)}$ for various values of $\alpha$ (the numbers at the end of the curves), for the quantum Ising model  \eqref{eq:Is} with $g=3/4$ and $\Delta=6$. At large time, the larger $\alpha$ the smaller the distance.
[\emph{Simulation parameters:} The initial time step is $\delta t_0=0.001$ and the cutoff norm is $10^{-7}$ the norm of the commutator with the Hamiltonian (which is inversely proportional to the time, see the inset of Fig. \ref{f:2}). Both parameters are updated dynamically to partially cope with the growth of complexity]}\label{f:0}
\end{figure}

Being only interested in the right hand side of \eqref{eq:exp},  the quality of the numerical procedure is essentially given by how well the conservation law $Q$ can be approximated by the time averaged quasi-conserved operators $\tilde Q^{(\alpha)}(t)$.  
In Fig. \ref{f:0}  the distance between the (opportunely normalised, \emph{cf.} \cite{SM}) time average of $G$ and $Q^{(\alpha)}$ is reported.  
Consistently with the assumption \eqref{eq:exp}, the distance decreases with time (for $\alpha>0$; notice that at time $t=0$ the distance from $Q^{(0)}\equiv G$ is zero) and saturates to a value that is smaller, the larger $\alpha$ is. 
We now turn to the issue of locality, using the time evolution to implicitly go to higher orders of perturbation theory.

Because of its complexity,  we can not use the operator norm to investigate the behaviour of the tails \eqref{eq:norm} and we must instead rely on some upper bound. An upper bound that can be readily computed with our algorithm is the \mbox{$L^1$-norm} of the coefficients of the expansion of the operator in involutions (in our case, strings of Pauli matrices): 
\be\label{eq:L1}
 \mathcal O=\sum_{j}\lambda_j  S_j\Rightarrow \parallel \mathcal O\parallel_1=\sum_j|\lambda_j|\qquad  S_j^2=\mathrm I\, .
\ee
We notice that the inequality $\parallel\mathcal O\parallel \leq \parallel\mathcal O\parallel_1$ can be saturated when the involutions associated with nonzero coefficients commute with each other.

\begin{figure}[tbp]
\includegraphics[width=0.45\textwidth]{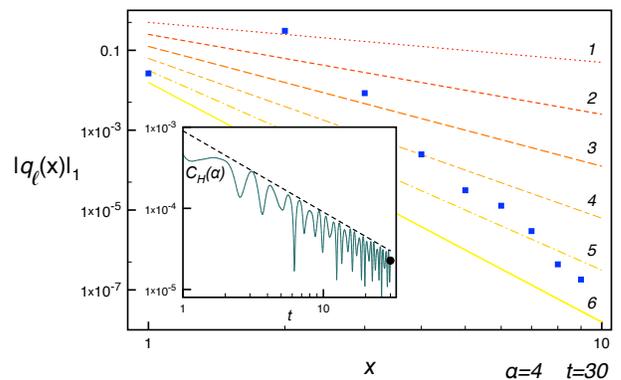}
\caption{The norm \eqref{eq:L1} of the operator density $q_\ell(x)$ (\emph{cf.} \eqref{eq:norm}) of $\tilde Q^{(4)}_{\rm HS}(t=30)$ as a function of the range $x$ (blue squares)  for the same system as in Fig. \ref{f:0}. Lines are guides to the eye: they have the slopes of different power laws with the exponent displayed right above the curves. 
The inset  can be used to estimate the relative precision, per unit of time, at which the operator is conserved:  
it shows $C_H(\alpha=4)=\parallel[H,\tilde Q^{(4)}_{\rm HS}]\parallel_{\rm HS}$ as a function of time. The dashed line is inversely proportional to the time. The black circle corresponds to $t=30$. We used the same simulation parameters as in Fig. \ref{f:0}.}\label{f:2}
\end{figure}

Figure \ref{f:2} shows the norm \eqref{eq:L1} of the operators that form $\tilde Q^{(4)}(t)$ as a function of their range.  At the time considered it is not clear whether the decay is exponential or algebraic.  In the latter case, 
 the tail could be a sign of the presence of other charges in \eqref{eq:exp}. In support of this conclusion we point out that the time evolution of $Q^{(\alpha)}$   for smaller values of $\alpha$ results in more pronounced tails (at fixed accuracy). 
 Nevertheless, the decay seems to be faster than a power law with an  exponent sufficiently large for $Q$ to ensure weak locality \eqref{eq:norm}.     
 In addition, a complementary analysis \cite{SM} in finite chains is compatible with $Q$ being quasi-local. 

Either quasi-locality or weak locality would have serious implications \emph{e.g.} on the relaxation properties in non-integrable models; consequently, this aspect demands extreme caution.   
We can not exclude the existence of a timescale dependent on $\Delta$ after which $\tilde Q^{(\alpha)}(t)$ will display some nonlocal character. Similarly, in finite chains the quasi-local behaviour could break down at larger sizes.
Nevertheless, $Q$ is a very good candidate for a quasi-local conservation law.

 \paragraph{Discussion.}
 We have constructed local quasi-conserved operators for a class of Hamiltonians that includes both integrable and non-integrable models. Their presence is expected to leave deep marks on the time evolution of local observables, which can remain frozen in a subspace of the Hilbert space\cite{quasi-stat, pre-therm,nonabelian, pre-thermLR}. This is particularly important in non-integrable models, in which the physical picture strongly relies on numerical studies. 
 
We conjectured that the quasi-conserved operators are an approximation of a conservation law and  investigated its locality properties.  
We found some numerical indications that the conservation law could be quasi-local, even in non-integrable models.  Our analysis is not conclusive but suggests that pre-thermalisation or even lack of thermalisation in non-integrable models\cite{strong-weakT,quasi-stat} could be understood in terms of a small number of (quasi-)local quasi-conserved operators, independent of the system details, and possibly being approximations of weakly local (\emph{cf.} \eqref{eq:norm}) conservation laws. 

Although we focussed on a particular class of models (\emph{cf.} \eqref{eq:H0}\eqref{eq:algebra}), our construction is easily generalisable, making us wonder whether analogous conservation laws are present in any spin-chain model with a local  Hamiltonian. 

Due to the limitations of the perturbative approach, only the strong coupling regime was investigated. To what extent this picture survives smaller values of the coupling constant is an important question to be considered in future research. 

\paragraph{Acknowledgments.} 
I thank Vincenzo Alba, Bruno Bertini, Pasquale Calabrese, Fabian Essler, and Neil Robinson for useful discussions. 
This work was supported by the EPSRC under grants EP/I032487/1 and EP/J014885/1.

\section*{Supplemental material}
\subsection{On the comparison of operators}
The numerical construction of a conservation law is affected by ambiguities that can depend on the procedure itself and make it impossibile to isolate a single charge. In the simplest situation the ambiguities can be parametrised by a few parameters. In a non-integrable model a minimal ansatz is given by \eqref{eq:exp}, which we rewrite here for the sake of clarity
\be\label{eq:expSM}
\tilde Q^{(\alpha)}(t)\xrightarrow{t\rightarrow \infty} \gamma  Q+\beta H \quad(+\hdots)\, ,
\ee 
where only the energy conservation is taken into account. 

In order to compare charges, we must first enforce the equivalence $\mathcal O+\beta H\sim \mathcal O$. We have chosen to redefine the operators in such a way to minimise the Hilbert-Schmidt norm $\parallel \mathcal O\parallel_{\rm HS}\propto\btr{}{\mathcal O^2}^{1/2}$.
Since
\be\label{eq:min}
0=\partial_\beta\parallel\mathcal O_0-\beta \mathcal O_1 \parallel_{\rm HS}\ \Rightarrow\  \beta=\frac{{\rm tr}[\mathcal O_0 \mathcal O_1]}{{\rm tr}[\mathcal O_1^2]}\, ,
\ee
the minimisation results in the redefinition 
\be\label{eq:shift}
\tilde Q^{(\alpha)}(t)\rightarrow \tilde Q^{(\alpha)}(t)-\frac{\btr{}{H\tilde Q^{(\alpha)}(t)}}{\btr{}{H^2}} H\, .
\ee

Concerning the normalisation, we prefer to keep the normalisation of $Q^{(\alpha)}$ unchanged.  
The normalisation of $\tilde Q^{(\alpha)}(t)$ can then be fixed by minimising the Hilbert-Schmidt distance between \eqref{eq:expSM} and $Q$. Since we do not know $Q$, 
in practice we minimise the distance with a quasi-conserved operator.  
For example, Fig. \ref{f:0} shows the time dependence of the following quantity (\emph{cf}. \eqref{eq:min}):
\be\label{eq:dHS}
d_{\rm HS}(Q^{(\alpha)},\tilde G(t))=\parallel Q^{(\alpha)}-\frac{\btr{}{Q^{(\alpha)}\tilde G(t)}}{\btr{}{(\bar G(t))^2}}\tilde G(t)\parallel_{\rm HS}\, ,
\ee  
where $Q^{(\alpha)}$ and $\tilde G(t)$ were redefined according to \eqref{eq:shift}.

\subsection{A non-perturbative conjecture}
The formal construction of the conservation law $Q$ relies on the non-trivial step of integrating the last equation of \eqref{eq:sys}. We propose a non-perturbative conjecture that allows us to calculate $\Omega_{n,0}$ recursively. 
Let 
\be\label{eq:int}
\mathcal K(x)=\mathrm T_\varphi\ e^{\smash{-i x \int_0^{2\pi}\mathrm d \varphi\,  e^{i \varphi}F +e^{-i \varphi} F^\dag -\sum\limits_{n>0,j}x^ne^{i j \varphi}\Omega_{n,j}}}\, ,
\ee 
where $\mathrm T_\varphi$ is the $\varphi$-ordering operator; we claim $\mathcal K(x)=\mathrm I$ for any $x$. This allows us to determine $\Omega_{n,0}$ unequivocally by series expanding $\mathcal K(x)$ about $x=0$ and imposing that the (operator) coefficients are zero (except for the trivial zeroth order). The meaning and the consequences of  $\mathcal K(x)=\mathrm I$ will be explored elsewhere. We just point out that, in most of the cases, this is equivalent to say that the spectrum of $H-Q$ is equally spaced.

\subsection{Exact diagonalisation}
Working out \eqref{eq:int} order by order can be rather involved, however the problem can in fact be overcome in finite chains. 

The first step is to construct a quasi-conserved approximation of the conserved operator. This can be done \emph{e.g.} using the perturbative results shown in Table \ref{t:2}.

The second step is to elevate the operator to an exactly conserved quantity. To that aim we project the quasi-conserved operator on the stationary states
\be
Q^{(\alpha)}\rightarrow \sum_{n}\braket{n|Q^{(\alpha)}|n}\ket{n}\bra{n}\, ,
\ee
where the sum is over an eigenbasis of the Hamiltonian and we assumed that $Q$ does not resolve the (exact) degeneracies of $H$. Although the final operator is exactly conserved, it is still only an approximation of $Q$, being the eigenvalues only approximately correct.

The next stage is genuinely non-perturbative: we correct the eigenvalues according to \eqref{eq:int}: 
\be\label{eq:n-p}
\braket{n|Q^{(\alpha)}|n}= Q_n+\delta Q^{(\alpha)}\xrightarrow{\eqref{eq:int}} Q_n\, .
\ee
This results in a conservation law that satisfies all our hypotheses: $[Q,H]=0$, $Q$ is approximated by $Q^{(\alpha)}$ at order $\mathcal O(\Delta^{-\alpha})$ and fulfils \eqref{eq:int}.

Having constructed $Q$, we can study the norm of the operators it consists of as a function of their range. 
In a finite chain the range can be defined as the minimal length of the connected subsystem in which the operator acts nontrivially.
The operators with fixed range $x$ smaller than $L/2+1$ can be singled out as follows:
\begin{multline}\label{eq:range}
q_\ell(x)=\frac{\mathrm{Tr}_{\overline{A}}[Q]}{\mathrm{Tr}[\mathrm I_{\overline{A}}]}\otimes \mathrm I_{\overline{A}}-\frac{\mathrm{Tr}_{\overline{A}\cup \mathcal R}[Q]}{\mathrm{Tr}[\mathrm I_{\overline{A}\cup \mathcal R}]}\otimes \mathrm I_{\overline{A}\cup \mathcal R}\\
-\frac{\mathrm{Tr}_{\mathcal L\cup \overline{A}}[Q]}{\mathrm{Tr}[\mathrm I_{\mathcal L\cup\overline{ A}}]}\otimes \mathrm I_{\mathcal L \cup\overline{ A}}+\frac{\mathrm{Tr}_{\mathcal L\cup \overline{A}\cup \mathcal R}[Q]}{\mathrm{Tr}[\mathrm I_{\mathcal L\cup \overline{A}\cup \mathcal R}]}\otimes \mathrm I_{\mathcal L \cup\overline{ A}\cup \mathcal R}\, ,
\end{multline}
where $|A|=x$, $\ell$ denotes the position of $A$, and we indicated with $\mathcal R$ and $\mathcal L$ the site at the right and at the left edge of $A$, respectively.  
The construction is a bit more involved when the range is larger than half a chain, including \eqref{eq:range} contributions already considered at smaller ranges, and being \eqref{eq:range} insensitive to the symmetries of the operators.

The relation between the range in the finite chain and the range in the thermodynamic limit is clearly ambiguous when the range is comparable with $L$; however, if the typical length of the operator is sufficiently smaller than the system size, the range in the finite chain is a good approximation of the one in the thermodynamic limit.

A first check of weak locality (or \emph{pseudo-locality} \cite{qlcons1}) is the behaviour of the operator norm of $Q$ per unit of length as a function of the length.
For $\Delta=2$  the asymptotic value is reached so quickly that the relative difference between $\parallel Q\parallel/L$ from $L=4$ to $L=12$ is smaller than $10^{-8}$. 

\begin{figure}[tbp]
\includegraphics[width=0.45\textwidth]{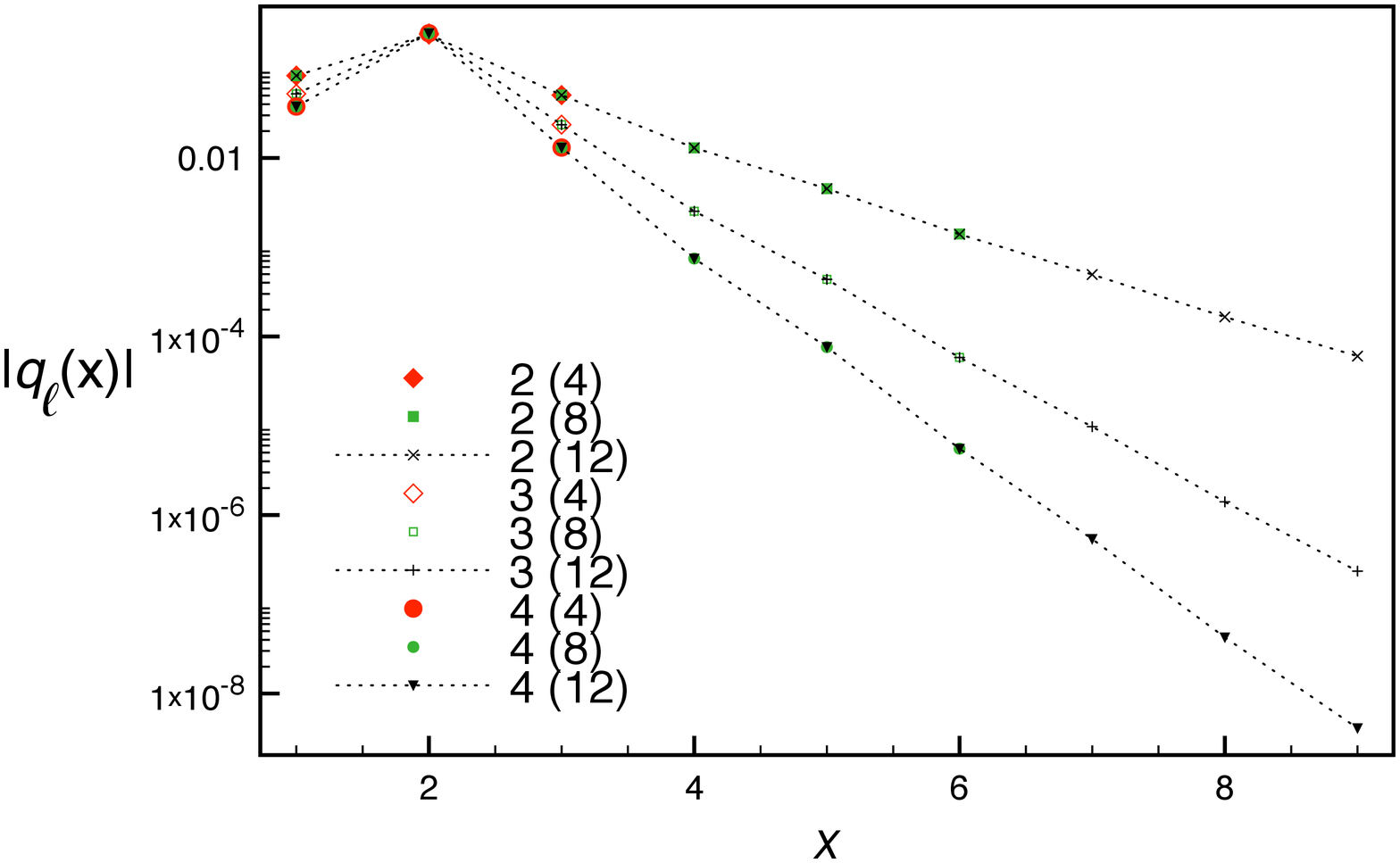}
\caption{The operator norm of the operator density $q_\ell(x)$ of $Q$ as a function of the range $x$ for the same system as in Fig. \ref{f:0} and various values of $\Delta$ and system sizes (in the legend $\Delta\ (L)$). The tail seems to decay exponentially.}\label{f:3}
\end{figure}

The same fast convergence is found in  $\parallel q_\ell(x)\parallel$ as a function of the range $x$. 
Fig. \ref{f:3} shows $\parallel q_\ell(x)\parallel$ for various values of $\Delta$ and system sizes. 
Our preliminary analysis in small chains is compatible with $Q$ being quasi-local. 
It should not be difficult to consider larger sizes, however the construction of $Q$ can become computationally demanding, also because, as the size is increased, the non-perturbative step requires a better quasi-conserved approximation of the conservation law.

Although this approach allows us to deal with the actual conservation law, it has an important weakness related to the order of limits. In the finite chain, because of the perturbative step, $\Delta$ turns out to be always comparable with $L$. On the other hand, we are interested in the limit $\Delta\ll L$, which is not easily accessible by exact diagonalisation techniques.

\end{document}